\documentclass[a4paper,12pt]{article}
\usepackage{amsmath}
\usepackage{amssymb}
\usepackage{textcomp}
\usepackage{graphicx}
\usepackage{bm}
\usepackage{color}
\newcommand{\beq}{\begin{equation}}
\newcommand{\eeq}{\end{equation}}
\newcommand{\myref}[1]{~{(\ref{#1})}}
\newcommand{\mycite}[1]{~{\cite{#1}}}

\newcommand{\be}{\begin{equation}}
\newcommand{\ee}{\end{equation}}
\newcommand{\ba}{\begin{eqnarray}}
\newcommand{\ea}{\end{eqnarray}}

\newcommand{\ns}{\nonumber\\}

\newcommand{\dmu}{\partial_{\mu}}

\newcommand{\dro}{\partial_{\rho}}
\newcommand{\dmuu}{\partial^{\mu}}
\newcommand{\dnuu}{\partial^{\nu}}
\newcommand{\dz}{\partial_{z}}

\newcommand{\lb}{\left(}
\newcommand{\rb}{\right)}
\newcommand{\lbf}{\left\{}
\newcommand{\rbf}{\right\}}
\newcommand{\ld}{\left.}

\newcommand{\rv}{\right|}
\newcommand{\lbr}{\left[}
\newcommand{\rbr}{\right]}

\begin{document}

\begin{titlepage}

\hfill\parbox{40mm}
{\begin{flushleft}  ITEP-TH-07/10
\end{flushleft}}

\vspace{10mm}

\begin{center}
{\large \bf On the Chiral Magnetic Effect in Soft-Wall AdS/QCD}

\vspace{17mm}

\textrm{A.~Gorsky$^1$, P.~N.~Kopnin$^{1,2}$ and
A.~V.~Zayakin$^{1,3}$} \vspace{8mm}

\textit{$^1$ Institute of Theoretical and Experimental Physics,\\
B.~Cheremushkinskaya ul. 25, 117259 Moscow, Russia}\\
\vspace{3mm}

\textit{$^2$ Moscow Institute of Physics and Technology,\\
Institutsky per. 9, 141 700 Dolgoprudny, Russia} \vspace{3mm}

\textit{$^3$ Fakult\"at f\"ur Physik der
Ludwig-Maximillians-Universit\"at M\"unchen und\\
 Maier-Leibniz-Laboratory, Am Coulombwall 1, 85748 Garching,
 Germany}

\vspace{3.5cm}

{\bf Abstract}\end{center} The essence of the chiral magnetic
effect is generation of an electric current along an external
magnetic field. Recently it has been studied by Rebhan \textit{et
al.} within the Sakai--Sugimoto model, where it was shown to be
zero. As an alternative, we calculate the chiral magnetic effect
in soft-wall AdS/QCD and find a non-zero result with the
natural boundary conditions. The mechanism of
the dynamical neutralization of the chiral chemical potential
via the string production is
discussed in the dual two-form representation.
\end{titlepage}

\nopagebreak[4]

\section{Introduction}
Chiral magnetic
effect\mycite{Alekseev:1998ds,Kharzeev:2007jp,Fukushima:2008xe}
(CME) is best described as a generation of an electric current by
a magnetic field in a topologically nontrivial background. The
standard field-theoretical argumentation is the following. Let us
consider QCD with massless quarks, so that left and right quarks
can be dealt with independently, and suppose a chiral chemical
potential $\mu_5$ is present, accounting for a certain
topologically nontrivial background. The topologically nontrivial
field configuration changes chirality, and an external magnetic
field $\mathbf{B}=(0,0,B)$ orders spins parallel to itself. Thus
arises a non-zero vector current, which is given by Fukushima,
Kharzeev and Warringa \cite{Fukushima:2008xe} \beq\label{kharzeev}
J_3^V=\frac{\mu_5 B}{2\pi^2} \equiv \mathcal{J}_{FKW}. \eeq

During recent years holography has become one of the main
alternative tools for analyzing non-perturbative QCD. Different
conductivities of quark matter, including chiral magnetic
conductivity, have already been analyzed in a variety of
holographic models. Electric conductivity in the D3/D7 model was
examined by Karch and O'Bannon in\mycite{Karch:2007pd}. Axial,
ohmic and Hall conductivity  in a magnetic field were calculated
on the basis of the Kubo formula and correlator analysis for the
Sakai-Sugimoto model in\mycite{Bergman:2008qv, Lifschytz:2009si}.
One of the results for electric current in\mycite{Bergman:2008qv}
is $1/2$ of QCD weak coupling result\myref{kharzeev}.

An attempt to describe the chiral magnetic effect for the vector
current in Sakai-Sugimoto model has been made recently in
\mycite{Yee:2009vw}. The result at zero frequency, where only the
Yang--Mills part of the action was used, exactly amounts to the
weak coupling QCD effect; non-zero frequencies have also been
considered. In\mycite{Rebhan:2009vc} a more sophisticated anomaly
subtraction scheme was suggested. It was argued that if one uses
the Bardeen term subtraction, then one gets zero effect for vector
current, otherwise one gets $2/3$ of the weak-coupling effect. The
reason for adding the Bardeen term to the action was to cure the
pathological behavior of the vector anomaly. Note that a similar
effect for the axial current in a theory with the conventional
chemical potential has been discussed in \cite{Metlitski:2005pr,
Newman:2005as}.

Experimental status of the problem is discussed in
\mycite{Voloshin:2009hr}. Presently it is claimed that the effect
is present, yet the exact proportionality coefficient $c$ in
$J_3^V=c\cdot\dfrac{\mu_5 B}{2\pi^2}$ cannot be inferred from it.
Lattice estimates are also close to
$2/3$\mycite{Buividovich:2009wi} of  weak-coupling effect. The
discussion of the effect in the framework of NJL model can be
found in \cite{Fukushima:2010fe}. Chiral magnetic effect at low
temperature was considered in~\cite{Nam:2009jb}. An analog of this
effect is known in superfluid helium~\cite{Volovik:2003fe}.

The present note aims at comparing the calculations of
\mycite{Rebhan:2009vc} to the chiral magnetic effect effect as
derived in the framework of soft-wall AdS/QCD.  The question
whether the effect is present in a holographic model or not, turns
out to be quite delicate. The paper is organized as follows. In
Section 2 we consider the analysis of the gauge sector of the
soft-wall model and confirm the result of \mycite{Rebhan:2009vc}.
In Section 3 we discuss the contribution of scalars and
pseudoscalars and focus on their boundary conditions in the 5d
equations of motion. Section 4 is devoted to the dual
representation of the chiral chemical potential and the mechanism
of its possible dynamical neutralization via the Schwinger-type
process. The results of this note are summarized in the
Conclusion.


\section{The soft-wall model}
\subsection{Gauge part of the action}

Let us consider the gauge field sector of the soft-wall AdS/QCD
model \cite{Karch:2006pv} taking into account the Chern-Simons
action $\int A\wedge F \wedge F$. We begin our consideration with
an action of Abelian fields L and R with a coupling $g_5$ that has
the following form: \ba S &=&
S_{YM}[L]+S_{YM}[R]+S_{CS}[L]-S_{CS}[R]\label{actionfull}\\
S_{YM}[A] &=& -\frac{1}{8g^2_5}\int e^{-\phi} F\wedge\ast F =
-\frac{1}{8g^2_5}\int dz\ d^4x\ e^{-\phi}\sqrt{g}
F_{MN}F^{MN}\label{actionYM}\\
S_{CS}[A] &=& -\frac{k\cdot N_c}{24\pi^2}\int A\wedge F\wedge F
-\frac{1}{2}A\wedge A\wedge A\wedge F +\frac{1}{10}A\wedge A\wedge
A\wedge A\wedge A \ns &=& -\frac{k\cdot N_c}{24\pi^2}\int dz\
d^4x\ \epsilon^{MNPQR}A_M F_{NP} F_{QR}.\label{actionCS}\ea

\noindent Here $k$ is an integer that scales the CS term and
effectively the magnetic field. Canonical normalization of the CS
term is $k=1$, but it will be kept it for the sake of generality.
The metric tensor is the following:

\beq ds^2 = g_{MN}dX^M dX^N = \dfrac{R^2}{z^2}\eta_{MN}dX^M dX^N =
\dfrac{R^2}{z^2}(-dz^2+dx_{\mu}dx^{\mu}). \eeq

\noindent In the $A_z=0$ gauge the YM action acquires the form \ba
S_{YM}[A] &=& -\frac{R}{4g^2_5}\int dz\ d^4x\ \left\{
\frac{e^{-\phi}}{z} A_{\mu}(\Box \eta^{\mu\nu} - \dmuu
\dnuu)A_{\nu} + A_{\mu} \dz \left( \frac{e^{-\phi}}{z}\ \dz
\right) A^{\mu} \right\}\ns &+& \left.\frac{R}{4g^2_5} \int d^4x\
\frac{e^{-\phi}}{z}A_{\mu}\dz A^{\mu}\right|_{z=0}^{z=\infty}
\label{actionYM1}. \ea

From the  YM part of the action we get \ba \frac{\delta
S_{YM}[A]}{\delta A_{\mu}} = -\frac{R}{2g_5^2}\lbf
\frac{e^{-\phi}}{z}(\Box \eta^{\mu\nu} - \dmuu \dnuu)A_{\nu} + \dz
\lb \frac{e^{-\phi}}{z} \dz A^{\mu} \rb \rbf .\label{dSYM}\ea

\noindent Varying the volume term of the action one gets \ba
\frac{\delta S_{CS}[A]}{\delta A_{\mu}} = \frac{k\cdot
N_c}{2\pi^2}\ \epsilon^{\mu\nu\rho\sigma}\ \dz A_{\nu}
F_{\rho\sigma}.\label{dSCS}\ea


Taking into account $\dfrac{R}{g^2_5} = \dfrac{N_c}{12\pi^2}$, one
obtains the equations of motion for the fields L and R \ba \dz \lb
\frac{e^{-\phi(z)}}{z} \dz L^{\mu} \rb - 24k
\epsilon^{\mu\nu\rho\sigma}\ \dz L_{\nu}
\partial_{\rho}L_{\sigma} = 0 \label{eomL}\\
\dz \lb \frac{e^{-\phi(z)}}{z} \dz R^{\mu} \rb + 24k
\epsilon^{\mu\nu\rho\sigma}\ \dz R_{\nu}
\partial_{\rho}R_{\sigma} = 0 \label{eomR}\ea
with the following boundary conditions

\begin{align} &L_0(0)=\mu_L, &R_0(0)&=\mu_R,\\
&L_3(0)=j_L, &R_3(0)&=j_R,\\
&L_1(0,x_2)=-\frac{1}{2}x_2B, &R_1(0,x_2)&=-\frac{1}{2}x_2B,\\
&L_{\mu}(\infty)=R_{\mu}(\infty), &\dz L_{\mu}(\infty)&=-\dz
R_{\mu}(\infty), \label{left=right}\end{align} here
$\mu=\dfrac{1}{2}(\mu_L+\mu_R), \mu_5=\dfrac{1}{2}(\mu_L-\mu_R)$,
and $j_{L,R}$ are the gauge field boundary values, a variation
with respect to which gives the currents \ba  \frac{\delta
S[L,R]}{\delta L_3(z=0)} = \frac{1}{V_{4D}}\frac{\partial
S[L,R]}{\partial j_L}
=\mathcal{J}_L,\\
\frac{\delta S[L,R]}{\delta R_3(z=0)} =
\frac{1}{V_{4D}}\frac{\partial S[L,R]}{\partial j_R}
=\mathcal{J}_R.\ea

\noindent Conditions (\ref{left=right}) arise because both left-
and right-handed gauge fields are associated with a single gauge
field in the Sakai--Sugimoto
model\mycite{SakaiSugimoto:2005le,SakaiSugimoto:2005mo}. In that
model, regions of positive and negative values of the holographic
coordinate $\rho=1/z$ correspond to left-handed $D8$ and
right-handed $\bar{D}8$ branes respectively. Since the gauge field
is smooth and continuous at $\rho=0$, a boundary condition
(\ref{left=right}) is obtained at $z=1/\rho = \infty$.
Furthermore, (\ref{left=right}) may be understood as a zero
boundary condition for the axial gauge field that is usually
imposed at the black hole horizon, which in our zero-temperature
case is located at $z=\infty$.

Denoting $\beta = 12kB$ one can get the following set of e.o.m.'s
\begin{align} &\dz \lb \frac{e^{-\phi(z)}}{z}\ \dz L_0 \rb = \beta \dz
L_3, &\dz \lb \frac{e^{-\phi(z)}}{z}\ \dz L_3 \rb &= \beta \dz
L_0,\\
&\dz \lb \frac{e^{-\phi(z)}}{z}\ \dz R_0 \rb = -\beta \dz R_3,
&\dz \lb \frac{e^{-\phi(z)}}{z}\ \dz R_3 \rb &= -\beta \dz
R_0,\\
&\dz \lb \frac{e^{-\phi(z)}}{z}\ \dz L_1 \rb = 0, &\dz \lb
\frac{e^{-\phi(z)}}{z}\ \dz R_1 \rb &= 0.\end{align}

\noindent Solution is the following \begin{align} &L_0(z) = \mu_L
+ \lb \mu_5 - \frac{1}{2}j_5 \rb \lb e^{-|\beta|w(z)}-1 \rb,
&L_3(z) =& j_L - \lb \mu_5 - \frac{1}{2}j_5 \rb \lb
e^{-|\beta|w(z)}-1 \rb,\ns &R_0(z) = \mu_R - \lb \mu_5 +
\frac{1}{2}j_5 \rb \lb e^{-|\beta|w(z)}-1 \rb, &R_3(z) =& j_R -
\lb \mu_5 + \frac{1}{2}j_5 \rb \lb e^{-|\beta|w(z)}-1 \rb,\ns
&R_1(z,x_2) = -\frac{1}{2}x_2B, &L_1(z,x_2) &=
-\frac{1}{2}x_2B,\label{solutions}\end{align} here $j=j_L+j_R,
j_5=j_L-j_R,$ and $w(z) = \int\limits_0^z du\ u\ e^{\phi(u)},\
\dfrac{e^{-\phi(z)}}{z}\ w'(z) = 1$.


\subsection{On-shell action and symmetry currents}

\noindent  Let us now compute the on-shell action with the gauge
fields given by (\ref{solutions}) for both left- and right-handed
gauge fields. Its Yang--Mills part is given as \ba S_{YM}[A] &=&
-\int dz\ d^4x\ \frac{e^{-\lambda z^2}}{z}\frac{R}{8g^2_5}\
\eta^{AB}\eta^{MN}F_{AM}F_{BN} \ns &=& -\frac{R}{8g^2_5} \int dz\
d^4x\ \frac{e^{-\lambda z^2}}{z} \lbf -2\eta^{\alpha\beta}\dz
A_{\alpha} \dz A_{\beta} +
\eta^{\alpha\beta}\eta^{\mu\nu}F_{\alpha\mu}F_{\beta\nu} \rbf \ns
&=& -\frac{R}{4g^2_5} \frac{B2}{4} V_{4D}\int dz\
\frac{e^{-\lambda z^2}}{z} \label{SYMcl} \ea The Chern--Simons
part of the action is \ba S_{CS}[A] &=& \frac{k\cdot
N_c}{6\pi^2}\int dz\ d^4x\ \epsilon^{\mu\nu\rho\sigma}\ A_{\mu}
\dz A_{\nu} F_{\rho\sigma} \ns &=& \frac{k\cdot N_c}{3\pi^2}\int
dz\ d^4x\ \lb A_0 \dz A_3 - A_3 \dz A_0 \rb F_{12}
.\label{SCScl}\ea

\noindent Recall that $ w(z) = \int\limits_0^z du\ u\ e^{\phi(u)},
w(0)=0, w(\infty)=\infty $. Then the solutions (\ref{solutions})
have the following form
\begin{align} &L_0(z) = \mu
+\frac{1}{2}j_5 + \lb \mu_5 -\frac{1}{2}j_5 \rb e^{-|\beta|w(z)},
&L_3(z) &= \mu_5 +\frac{1}{2}j - \lb \mu_5 -\frac{1}{2}j_5 \rb
e^{-|\beta|w(z)},\ns &R_0(z) = \mu +\frac{1}{2}j_5 - \lb \mu_5
+\frac{1}{2}j_5 \rb e^{-|\beta|w(z)}, &R_3(z) &= \mu_5
+\frac{1}{2}j - \lb \mu_5 +\frac{1}{2}j_5 \rb e^{-|\beta|w(z)},\ns
&F^L_{12} = \frac{1}{2}B, &F^R_{12} &= \frac{1}{2}B.
\label{solutions1}\end{align}

\noindent Upon substituting (\ref{solutions1}) into (\ref{SCScl})
the on-shell CS action becomes \ba S_{CS}[L] &=& \frac{k\cdot
N_c}{3\pi^2}\int dz\ d^4x\ F^L_{12}\lb L_0\dz L_3 - L_3\dz L_0 \rb
\ns &=& \frac{k\cdot N_c}{6\pi^2} B V_{4D} \lb \mu\mu_5 + \mu_5^2
- \frac{1}{2}\mu j_5 + \frac{1}{2}\mu_5 j - \frac{1}{4}j_5^2 -
\frac{1}{4}j j_5 \rb,\label{SCSLcl} \ea

\noindent and \ba S_{CS}[R] &=& \frac{k\cdot N_c}{3\pi^2}\int dz\
d^4x\ F^R_{12}\lb R_0\dz R_3 - R_3\dz R_0 \rb \ns  &=&
\frac{k\cdot N_c}{6\pi^2} B V_{4D} \lb \mu\mu_5 - \mu_5^2 +
\frac{1}{2}\mu j_5 - \frac{1}{2}\mu_5 j + \frac{1}{4}j_5^2 -
\frac{1}{4}j j_5 \rb.\label{SCSRcl} \ea

\noindent The symmetry currents $\mathcal{J}_L, \mathcal{J}_R$ are
equal to the partial derivatives of the action with respect to
$j_L, j_R$ \ba \mathcal{J}_L &=& \frac{1}{V_{4D}} \frac{\partial
S}{\partial j_L} = \frac{1}{V_{4D}} \lb \frac{\partial j}{\partial
j_L} \frac{\partial S}{\partial j} + \frac{\partial j_5}{\partial
j_L} \frac{\partial S}{\partial j_5} \rb = \frac{1}{V_{4D}} \lb
\frac{\partial S}{\partial j} + \frac{\partial S}{\partial j_5}
\rb,\\
\mathcal{J}_R &=& \frac{1}{V_{4D}} \frac{\partial S}{\partial j_R}
= \frac{1}{V_{4D}} \lb \frac{\partial j}{\partial j_R}
\frac{\partial S}{\partial j} + \frac{\partial j_5}{\partial j_R}
\frac{\partial S}{\partial j_5} \rb = \frac{1}{V_{4D}} \lb
\frac{\partial S}{\partial j} - \frac{\partial S}{\partial j_5}
\rb,\\
\mathcal{J} &=& \frac{2}{V_{4D}} \frac{\partial S}{\partial j}\ ,\
\mathcal{J}_5 = \frac{2}{V_{4D}} \frac{\partial S}{\partial j_5}\
.\label{J}\ea

\noindent As can be seen from (\ref{SYMcl}), the YM part of the
action does not depend on the current sources. The CS part, on the
other hand, equals \ba S_{CS} = S_{CS}[L] - S_{CS}[R] =
\frac{k\cdot N_c}{6\pi^2} B V_{4D} \lb 2\mu_5^2 - \frac{1}{2}j_5^2
+ \mu_5 j - \mu j_5 \rb. \label{Sfinal}\ea

\noindent From eqs. (\ref{J},\ref{Sfinal}) one obtains
\ba \mathcal{J} &=& \frac{k\cdot N_c}{3\pi^2} B \mu_5,\label{current}\\
\mathcal{J}_5 &=& -\frac{k\cdot N_c}{3\pi^2} B (\mu + j_5).
\label{axialcurrent}\ea

\noindent If one sets $k=1$, a standard normalization of the CS
action (\ref{actionCS}) is recovered and the result
(\ref{current}) is in agreement with\mycite{Rebhan:2009vc} without
the Bardeen counterterm. The axial supercurrent introduced
in\mycite{Rebhan:2009vc} is an equivalent of our $j_5$. If it is
interpreted as a source for the axial current it has to be set to
zero. Minimizing the action with respect to it is analogous to
setting the axial current (\ref{axialcurrent}) to zero. It is
interesting that the answer does not depend on $j$, which probably
justifies its absence in\mycite{Rebhan:2009vc}.


\subsection{The divergence of the vector current}

\noindent In this section the general formula for the left and
right symmetry currents $\mathcal{J}_{L,R}$ will be derived.

\noindent  We are going to use a different approach to calculating
currents, yet the results will be identical to the previous
section. The current definition is the following (for the current
$\mathcal{J}_{L,R}\equiv \mathcal{J}_{L,R}^3$ it is the same as in
(\ref{J})): \ba \mathcal{J}^{\mu}_{L}(x) = \frac{\delta S}{\delta
L_{\mu}(z=0,x)},\ \mathcal{J}^{\mu}_{R}(x) = \frac{\delta
S}{\delta R_{\mu}(z=0,x)}. \ea

\noindent The variation of the action $\delta S=\delta
S_{YM}+\delta S_{CS}$ can be split into two parts -- one
proportional to the equations of motion and one reducible to a
surface term \ba \delta S_{YM}[A] &=& \delta S^{vol}_{YM}[A] - \ld
\frac{R}{2g^2_5}\int d^4x \frac{e^{-\phi(z)}}{z}\ \dz A^{\mu}
\delta A_{\mu} \rv_{z=0},\label{varYM}\\
\delta S_{CS}[A] &=& \delta S^{vol}_{CS}[A] + \ld\frac{k\cdot
N_c}{6\pi^2}\int d^4x\
\epsilon^{\mu\nu\rho\sigma}A_{\nu}F_{\rho\sigma}\delta
A_{\mu}\rv_{z=0}.\label{varCS}\ea

\noindent The 5D parts are equal to zero on-shell, so that one
gets \ba \mathcal{J}^{\mu}_{L}(x) &=& -\frac{R}{2g^2_5}
\frac{e^{-\phi(z)}}{z}\ \dz L^{\mu} + \frac{k\cdot N_c}{6\pi^2}\
\epsilon^{\mu\nu\rho\sigma}\ L_{\nu}F^L_{\rho\sigma},\label{JL}\\
\mathcal{J}^{\mu}_{R}(x) &=& -\frac{R}{2g^2_5}
\frac{e^{-\phi(z)}}{z}\ \dz R^{\mu} - \frac{k\cdot N_c}{6\pi^2}\
\epsilon^{\mu\nu\rho\sigma}\ R_{\nu}F^R_{\rho\sigma}.\label{JR}
\ea

\noindent Recalling that $\dfrac{R}{g^2_5}=\dfrac{N_c}{12\pi^2}$
and $V_{\mu} = L_{\mu}+R_{\mu}$  the following expression for the
divergence of the vector current is obtained \ba \dmu
\mathcal{J}^{\mu} &=& \dmu
(\mathcal{J}^{\mu}_{L}+\mathcal{J}^{\mu}_{R}) =
-\frac{N_c}{24\pi^2}\frac{e^{-\phi(z)}}{z}\ \dz\dmu V^{\mu} +
\frac{k\cdot N_c}{6\pi^2}\ \epsilon^{\mu\nu\rho\sigma}\ (\dmu
L_{\nu}F^L_{\rho\sigma}-\dmu R_{\nu}F^R_{\rho\sigma})\ns &=&
-\frac{N_c}{24\pi^2}\frac{e^{-\phi(z)}}{z}\ \dz\dmu V^{\mu} +
\frac{k\cdot N_c}{3\pi^2}\ \epsilon^{\mu\nu\rho\sigma}\ \dmu
V_{\nu}\dro A_{\sigma}.\label{div1}\ea

\noindent To express the divergence of the vector field $V_{\mu}$
another equation of motion generated by $\dfrac{\delta S}{\delta
A_z}$ will be needed \ba \frac{\delta S_{YM}[A]}{\delta A_z} &=&
\frac{R}{2g^2_5}\frac{e^{-\phi(z)}}{z}\ \dz\dmu A^{\mu} =
\frac{N_c}{24\pi^2}\frac{e^{-\phi(z)}}{z}\ \dz\dmu A^{\mu},\ns
\frac{\delta S_{CS}[A]}{\delta A_z} &=&
-\frac{kN_c}{24\pi^2}\frac{\delta}{\delta A_z}\int d^4x\ dz\
\epsilon^{\mu\nu\rho\sigma}(A_z F_{\mu\nu}F_{\rho\sigma} -
4A_{\mu}F_{z\nu}F_{\rho\sigma})=\ns &=& -\frac{kN_c}{2\pi^2}\
\epsilon^{\mu\nu\rho\sigma}\ \dmu A_{\nu} \dro
A_{\sigma}.\label{eomz}\ea

\noindent The corresponding e.o.m.'s assume the form: \ba
\frac{e^{-\phi(z)}}{z}\ \dz\dmu L^{\mu} = 12k
\epsilon^{\mu\nu\rho\sigma}\ \dmu L_{\nu} \dro L_{\sigma},&&
\frac{e^{-\phi(z)}}{z}\ \dz\dmu R^{\mu} = 12k
\epsilon^{\mu\nu\rho\sigma}\ \dmu R_{\nu} \dro R_{\sigma},\ns
\frac{e^{-\phi(z)}}{z}\ \dz\dmu V^{\mu} &=& 12k
\epsilon^{\mu\nu\rho\sigma}\ \dmu V_{\nu} \dro A_{\sigma}.
\label{eomzlrv}\ea

\noindent Thus the divergence in (\ref{div1}) equals: \ba \dmu
\mathcal{J}^{\mu} &=& -\frac{kN_c}{6\pi^2}\
\epsilon^{\mu\nu\rho\sigma}\ \dmu V_{\nu} \dro
A_{\sigma}.\label{div2}\ea


\subsection{The Bardeen counterterm}

The Bardeen counterterm has a dimensionless prefactor $c$ that is
determined by the following condition \be \dmu
\mathcal{J}_{subtracted}^{\mu} = \dmu \mathcal{J}^{\mu} + \dmu
\mathcal{J}_{Bardeen}^{\mu} = 0,\label{bardeen}\ee \noindent
where the counterterm has the form
 \be S_{Bardeen} = c \int
d^4x
\epsilon^{\mu\nu\rho\sigma}L_{\mu}R_{\nu}(F^L_{\rho\sigma}
+F^R_{\rho\sigma}).\label{Sbardeen1}
\ee
 It may be interpreted as a surface counterterm in the spirit of
 the holographic renormalization. In our case
 \ba S_{Bardeen} &=& -2c \ld \int d^4x (L_0R_3-L_3R_0)\partial_2 V_1 \rv_{z=0} =
 2cBV_{4D}\ld(L_0R_3-L_3R_0)\rv_{z=0}=\ns &=&
 2cBV_{4D}(\mu_Lj_R-\mu_Rj_L) = 2cBV_{4D}(\mu_5 j-\mu j_5).
 \label{Sbardeen2}\ea

 \noindent  hence
 \be \mathcal{J}_{Bardeen} = 4cB\mu_5,\,\, \mathcal{J}_{Bardeen\ 5} = -4cB\mu .\label{Jbardeen}\ee

 \noindent The general expression for the currents is the following:
 \ba \delta S_{Bardeen} &=& \int d^4x \left(
 \mathcal{J}^{\mu}_{Bardeen\ L}\ \delta L_{\mu}
 + \mathcal{J}^{\mu}_{Bardeen\ R}\ \delta
 R_{\mu},\right)\ns \mathcal{J}^{\mu}_{Bardeen\ L} &=&
 2c\epsilon^{\mu\nu\rho\sigma}\lb R_{\nu}\dro R_{\sigma}+2R_{\nu}\dro L_{\sigma}+L_{\sigma}\dro R_{\nu}
 \rb, \ns \mathcal{J}^{\mu}_{Bardeen\ R} &=&
 -2c\epsilon^{\mu\nu\rho\sigma}\lb L_{\nu}\dro L_{\sigma}+2L_{\nu}\dro R_{\sigma}+R_{\sigma}\dro L_{\nu}
 \rb, \ns \mathcal{J}^{\mu}_{Bardeen} &=&
 2c\epsilon^{\mu\nu\rho\sigma}\lb R_{\nu}\dro R_{\sigma} - L_{\nu}\dro L_{\sigma} - 3L_{\nu}\dro R_{\sigma} + 3R_{\nu}\dro L_{\sigma}
 \rb
.\label{Jbardeenfull}\ea

\noindent The divergence of the Bardeen current equals: \be \dmu
\mathcal{J}^{\mu}_{Bardeen} = -2c\epsilon^{\mu\nu\rho\sigma}\dmu
V_{\nu}\dro A_{\sigma}\label{divbardeen} \ee

\noindent  Based on (\ref{div2}, \ref{bardeen}, \ref{divbardeen})
one gets \be c=-\frac{kN_c}{12\pi^2}.\label{constbardeen} \ee

\noindent  As a result, the subtracted current turns out to be
zero (\ref{current}, \ref{Jbardeen}, \ref{constbardeen}): \be
\mathcal{J}_{subtracted} = \mathcal{J}+\mathcal{J}_{Bardeen} =
\frac{kN_c}{3\pi^2}B\mu_5 + 4cB\mu_5 = \frac{kN_c}{3\pi^2}B\mu_5 +
4\lb-\frac{kN_c}{12\pi^2}\rb B\mu_5 = 0.\label{Jfull}\ee


\section{Scalars and Pseudoscalars}
\subsection{Kinetic term and potential}
Let us consider now the scalar--pseudoscalar sector, which was
first omitted from our considerations. The bilinear part is \beq
S_X=\int d^4x dz\ e^{-\phi}\ R^3 \left[\frac{1}{z^3}(D^\mu
X)^\dagger D^\mu X + \frac{3}{z^5}|X|^2\right] \eeq where
$D_\mu=\partial_\mu X-i L_\mu X +iX R_\mu$; field $X$ is related
to pion field via \beq\label{pion} X=\exp{\left(i\frac{\pi^a
t^a}{f_\pi}\right)}\frac{1}{2}v(z). \eeq What is crucial for our
case is that there is a scalar interaction with gauge fields. It
can be thought of in two different ways. If one works in $A_z=0$
gauge (both $L_z=0$ and $R_z=0$), than pion is identified with
phase of field $X$ as in\myref{pion}, and interaction term is
\beq\label{XAA} S_{XAA}=\frac{N_c}{24\pi^2}\ \mathrm{tr}\int d^4x
dz\
\partial_{\mu}V_\nu
\partial_{\lambda}V_\rho \frac{\partial_\alpha \pi}{f_\pi}\ \epsilon^{\mu\nu\lambda\rho\alpha}.
\eeq If, however, one does not impose this gauge, then the
holonomy of the axial field $\int A_z dz$ is itself the pion
field, and the term\myref{XAA} arises directly form Chern--Simons.
Note there is no double-counting here: when dealing with
Chern-Simons solely (as was the case in the Sakai-Sugimoto model
of\mycite{Rebhan:2009vc}),  $A_z$ can always be set to zero. This
is impossible without inducing a phase of $X$ in the true AdS/QCD
model by Karch--Katz--Son--Stephanov\mycite{Karch:2006pv} which we
work in.

Taking action\myref{XAA} and differentiating it over $F_{z3}$ one
gets the following contribution to current \beq
\mathcal{J}_{XAA}=\frac{N_c}{2\pi^2}\frac{1}{3}B\dfrac{\partial_0
\pi(x)}{f_{\pi}}. \eeq

The special care concerns the boundary conditions. We would like
to argue that the linear in time ``rotating"  boundary conditions
are appropriate. Let us remind the PCAC relation connecting the
axial current and the pion field \beq
\bar{\Psi}\gamma_{\nu}\gamma_5\Psi \Leftrightarrow
f_{\pi}\partial_{\nu}\pi \eeq which implies that we add the
following term in the pion lagrangian \beq \mu_5
f_{\pi}\partial_{0}\pi. \eeq This term changes the pion canonical
momentum and condition of the vanishing of the canonical momentum
yields the rotating boundary condition \beq P=\partial_{0}\pi +
\mu_5 f_{\pi}= 0 \eeq

Collecting all the terms we get
\beq
\begin{array}{l}
\mathcal{J}_{full,\ subtracted}=\mathcal{J}+\mathcal{J}_{Bardeen}+\mathcal{J}_{XAA}=\frac{1}{3} \mathcal{J}_{FKW}\\
\mathcal{J}_{full,\
nonsubtracted}=\mathcal{J}+\mathcal{J}_{XAA}=\mathcal{J}_{FKW}
\end{array}
\eeq


\subsection{Chern-Simons action with scalars}

The result of the previous section can be justified from a
somewhat different point of view. Let us once more consider the
Chern--Simons action

\be S_{CS} = \frac{-kN_c}{24\pi^2}\lb \int L \wedge dL \wedge dL -
\int R \wedge dR \wedge dR \rb.\ee

\noindent Its gauge transformation ($L \rightarrow L+d\alpha_L$,
$R \rightarrow R+d\alpha_R$) is proportional to a surface term

\be S_{CS} \rightarrow S_{CS} + \frac{-kN_c}{24\pi^2}\lb \int
d\alpha_L \wedge dL \wedge dL - \int d\alpha_R \wedge dR \wedge dR
\rb .\ee

\noindent While in the standard field theory this is satisfactory,
in our particular consideration this boundary term is nonzero and
the gauge invariance is violated.  When the
component $A_z$ is gauged out, one has to introduce in some other
way the pion back into the Chern--Simons action.

One may proceed in the following way. An explicitly gauge
invariant Chern--Simons term is defined as

\be \bar{S}_{CS} = \frac{-kN_c}{24\pi^2}\lb \int (L+d\phi_L)
\wedge dL \wedge dL - \int (R+d\phi_R) \wedge dR \wedge dR \rb,\ee
where $\phi_{L,R}$ are scalar fields that transform so as to keep
the combinations within the brackets invariant, $\phi_{L,R}
\rightarrow \phi_{L,R} - \alpha_{L,R}$. This means that the
combination $f_{\pi}\lb\phi_R - \phi_L\rb$ may be associated with
the five-dimensional pion in the gauge in which $A_z$ is set to
zero.

As in the previous section arguments can be made in favor of
setting the scalar fields proportional to the chemical potentials
on the ultraviolet holographic boundary at $z=0$

\be \ld\phi_{L,R}\rv_{z=0} = -\frac{1}{2}\mu_{L,R}\cdot t .\ee

To clarify the infrared behavior ($z \rightarrow \infty$) of the
scalars let us turn once more to the Sakai--Sugimoto
model\mycite{SakaiSugimoto:2005le,SakaiSugimoto:2005mo}, in which
the infrared region is the area where the $D8$ and $\bar{D}8$
branes connect. There the Chern--Simons action is a single
integral over both $D8$ and $\bar{D}8$ branes

\be S_{CS} \sim \int A \wedge dA \wedge dA ,\ee where the
holographic coordinate $\rho=1/z$ runs from $-\infty$ (right
$\bar{D}8$ brane) to $\infty$ (left $D8$ brane) and the gauge
field $A$ is associated with the left-handed field $L$ of the
Karch--Katz--Son--Stephanov model at $\rho>0$ and with the
right-handed field $R$ at $\rho<0$.

We might undertake a similar procedure of making this action
explicitly gauge invariant by adding a single scalar $\phi$

\noindent \be \bar{S}_{CS} \sim \int (A+d\phi) \wedge dA \wedge dA
,\ee \noindent and this scalar field will be analogous to $\phi_L$
($\phi_R$) for positive (negative) values of $\rho$.

Now, since the field $\phi$ is smooth and continuous at $\rho=0$ a
boundary condition is obtained for the $\phi_{L,R}$ fields in our
setup

\be \mbox{for all } x_{\mu} \ld \phi_L \rv_{z=\infty}= \ld \phi_R
\rv_{z=\infty} .\label{bcphi}\ee \noindent It is quite analogous
to the condition (\ref{left=right}) from the third section of this
paper.

In what follows it will be assumed that the gauge fields are
adiabatically tuned out in the temporal positive and negative
infinities.

Let us simplify the modification of the Chern--Simons action
$S_{\phi A A}=\bar{S}_{CS}-S_{CS}$ which happens to be a surface
term

\ba S_{\phi A A}&=& \frac{-kN_c}{24\pi^2}\lb \int d\phi_L \wedge
dL \wedge dL - \int d\phi_R \wedge dR \wedge dR \rb \ns &=&
\frac{kN_c}{6\pi^2}B\lb \int dz\ d^4x\ \partial_t \phi_L \dz L_3 -
\int dz\ d^4x\ \partial_t \phi_R \dz R_3 \rb .\ea

\noindent If the surface terms that arise at temporal infinities
$\sim \ld \int dz\ d^3x \{L,R\}_3 \dz
\phi_{L,R}\rv^{t=+\infty}_{t=-\infty}$ are neglected, the
following is obtained

\ba S_{\phi A A}&=& \frac{kN_c}{6\pi^2}B\lbr -
\vphantom{\frac{1}{1}} \lb j_L\partial_t \phi_L - j_R\partial_t
\phi_R \rb\rv_{z=0} \ns &+& \ld\ld\lb \mu_5 + \frac{1}{2}j_L +
\frac{1}{2}j_R\rb \lb
\partial_t \phi_L - \partial_t \phi_R \rb \rv_{z=\infty} \rbr .\ea

\noindent Due to the boundary condition (\ref{bcphi}) the second
term vanishes and another contribution to the current is found

\be \mathcal{J}_{\phi A A} = \frac{kN_c}{6\pi^2}B\mu_5
\label{jphiAA}.\ee

\noindent The total current now equals

\ba \mathcal{J}_{full,\ subtracted} &=&
\mathcal{J}+\mathcal{J}_{Bardeen}+\mathcal{J}_{\phi A A} =
\frac{1}{3}\frac{kN_c}{2\pi^2}B\mu_5 .\label{Jfullrenorm}\\
\mathcal{J}_{full,\ nonsubtracted} &=&
\mathcal{J}+\mathcal{J}_{\phi A A} = \frac{kN_c}{2\pi^2}B\mu_5
.\label{Jfullnonrenorm} \ea

\noindent Here is a summary of all the contributions to the chiral
magnetic effect
(\ref{J},\ref{Jbardeen},\ref{Jbardeenfull},\ref{Jfull},\ref{jphiAA},\ref{Jfullnonrenorm},\ref{Jfullrenorm}).
\vskip1cm \noindent
\begin{tabular}{|c|c|c|c|c|c|c|}
\hline Term in  & \multicolumn{2}{|c|}{Yang--Mills} &
\multicolumn{2}{|c|}{Chern--Simons} & Bardeen
 & Scalars \\
\cline{2-3} \cline{4-5}
the action & bulk & boundary & bulk & boundary & counterterm & in CS\\
\hline &&&&&&\\[5pt]
Contribution & $-\dfrac{1}{3}\dfrac{N_c}{2\pi^2}B\mu_5$ &
$\dfrac{1}{3}\dfrac{N_c}{2\pi^2}B\mu_5$ &
$\dfrac{1}{3}\dfrac{N_c}{2\pi^2}B\mu_5$ &
$\dfrac{1}{3}\dfrac{N_c}{2\pi^2}B\mu_5$ &
$-\dfrac{2}{3}\dfrac{N_c}{2\pi^2}B\mu_5$ &
$\dfrac{1}{3}\dfrac{N_c}{2\pi^2}B\mu_5$\\
to the current &&&&&&\\[5pt] \hline
\end{tabular}

\vskip1cm \noindent
\begin{tabular}{|c|c|c|c|c|}
\hline
Action taken & \multicolumn{2}{|c|}{Total} & \multicolumn{2}{|c|}{Total without scalars} \\
\cline{2-5}
into account & subtracted & nonsubtracted & subtracted & nonsubtracted\\
\hline Resulting current, &&&&\\[5pt]
in terms of $\dfrac{N_c}{2\pi^2}B\mu_5$ & $\dfrac{1}{3}$ & $1$ &
$0$ &
$\dfrac{2}{3}$\\
 &&&&\\[5pt] \hline
\end{tabular}

That is, we see that with the Bardeen counterterm excluded we have
reproduced the initial result for the CME current. The natural
question concerns the relation of our answer with the one obtained
in the Sakai-Sugimoto model. Naively both models describe the same
system, however in the soft-wall model we have extracted the
contribution from the scalars explicitly. The point which yields
the different answers follows from the nontrivial boundary
conditions imposed on the meson fields in the  CME situation. If
we take them carefully into account, the additional term has to be
considered in the Sakai-Sugimoto model as well.  Note that the
question concerning the necessity of Bardeen counterterm is very
subtle and deserves the additional study. It is the account of
this term that results in an answer for the axial current in a
theory with a vector chemical potential which is different from
\cite{Metlitski:2005pr,Newman:2005as}.

\section{On the dynamical neutralization of $\mu_5$}
In this Section we make a comment concerning the dual
interpretation of the chiral chemical potential and mention the
nonperturbative effect of its dynamical neutralization most
clearly seen in the dual formulation. Let us first remind the
expression for the Goldstone-Wilczek current \cite{gw} for the
fermions in the external pseudoscalar and electromagnetic fields
\beq J_{GW,\mu}=\epsilon_{\mu\nu\alpha\beta} \partial_{\nu}\phi
F_{\alpha\beta} \eeq That is, the chiral magnetic effect in a
generic situation can emerge from the pseudoscalar field linear in
time, $\phi = \mu_5 t$.

Consider now the dual representation of the pseudoscalar field
in $D=4$ which reads as
\beq
\partial_{\mu} \phi \propto \epsilon_{\mu\nu\alpha\beta}\partial_{\nu}B_{\alpha\beta}
\eeq\noindent  where $B_{\mu\nu}$ is the antisymmetric two-form
field which in $D=4$ has only one nontrivial degree of freedom. The
non-vanishing constant $\mu_5$ means that we have constant
curvature of the two-form field $H=dB\propto \mu_5$. Actually $\mu_5$
corresponds to the "magnetic" part of the curvature H.

The string is charged with respect to the two-form field, that is,
at $\mu_5\neq 0$ the issue of magnetic string production via
Schwinger mechanism can be posed. This process of closed string
production screens the initial value of $\mu_5$. To calculate the
probability rate one can consider the Euclidean classical solution
for the string in the external field. The classical solution
corresponds to the spherical worldsheet and the surface tension
term is compensated by the volume term from the external field
\beq S_{eff}(R)= \pi R^2 T - 4/3\ \pi R^3 H \eeq where T is the
effective tension of the string. The extremization of the
effective action yields the critical radius R of the bounce
configuration \beq R_0=\frac{T}{2H}. \eeq

\noindent That is, the non-perturbative probability rate to create
the closed string reads as \beq w\propto \exp {\left(-\frac{\pi
T^3}{12H^2}\right)} \eeq \noindent To justify the validity of the
semi-classical analysis the action calculated at the bounce
configuration is assumed to be large. To this aim let us note that
the magnetic string  can be considered as the D2 brane wrapped
over one internal direction \cite{Gorsky:2009me} and its tension
tends to vanish at the $T<T_c$. That is at small temperatures the
situation is similar to the absolute  instability of the electric
field in the massless QED. In the case under consideration the
$\mu_5$ background at $T<T_c$ is similarly unstable with respect
to the magnetic string production and the very introduction of the
chiral chemical potential seems to be impossible.

However, it is reasonable to discuss this process above the phase
transition. Two features of this region are relevant. First, the
tension of the magnetic string above the transition point is
finite \cite{Gorsky:2009me} and the semi-classical calculation
above is reasonable and the notion of the chiral chemical
potential can be introduced. Secondly, the problem becomes
effectively three-dimensional and the two-form field in $D=3$
becomes non-dynamical. This means that the value of the
pseudoscalar can only jump crossing the string worldsheets. It
provides a kind of capacitor for the strings.

Usually the tunneling is assumed to proceed at zero energy,
however having in mind that in the heavy-ion collisions the
induced tunneling at nonzero energy has to be considered at equal
footing. There is some peculiarity concerning the induced string
production in the external two-form field. The point is that the
process involves the intermediate Minkowski region before the
Euclidean part of the path similar to the induced false vacuum
decay in the $D>2$. The Minkowski region solution  corresponds to
the oscillating string which at the transition point has to be
glued with the Euclidean solution. Such a two-step solution has
been considered in \cite{gv} in a slightly different context. Let
us also note that at some quantized values of the energy the
resonant tunneling takes place and the probability rate for the
string production increases dramatically. The example of the
resonant tunneling has been considered in \cite{gv}.

It was suggested in \cite{Kharzeev:2007jp} that non-vanishing value
of the chiral chemical potential is due to the instantons. The
discussion of this section implies that the instantons are not
directly related to $\mu_5$ since they do not provide the constant
curvature of the two-form field and the different nonperturbative
solutions like magnetic string could be relevant. The role of the instantons
in the two-form field theory is different and they yield the effective
mass for the fluctuations of the two-form field similar to the
generation of the axion mass in the dual representation \cite{sus}.

\section{Conclusion}
In this note  the holographic derivation of the chiral magnetic
effect has been revisited in the soft-wall AdS/QCD model. Unlike
the estimate via the Sakai-Sugimoto model \cite{Rebhan:2009vc}, in
soft-wall model the effect is present under reasonable boundary
conditions; the difference between our model and one in
\cite{Rebhan:2009vc} is the presence of an additional contribution
from the scalar part of the action. Putting it loosely, scalars
act as ``catalysts'' for the effect, the value of which is
determined however not by those, but by the Chern--Simons term.
Thus the effect is still topological in its nature, as it is
within the standard paradigm; though triggered by scalars, it is a
robust prediction in the sense of its independence on the
Lagrangian of the scalars. Notice that the effect is trivially
absent in the D3/D7 model due to the different form of the
Chern-Simons term.

The dual representation of the chiral chemical potential has been
suggested and an effect which depends on $\mu_5$ non-analytically
has been mentioned. This non-perturbative Schwinger-type effect of
string creation is responsible for the dynamical neutralization of
the chiral chemical potential and we have emphasized its strong
dependence on the temperature.

We are grateful to O.~Bergman, L.~Brits, D.~Kharzeev,
K.~Landsteiner, G.~Lifschytz, A.~Krikun, V.~Shevchenko and
V.~Zakharov for useful discussions. A.Z. specially thanks
J.~Erdmenger and D.~Habs for their advice.
This work was supported in part by the RFBR grant RFBR 10-01-00836
(A.Z.), PICS-07-0292165 (A.G.), RFBR-09-02-00308 (A.G., P.K.),
CRDF - RUP2-2961-MO-09 (A.G.) and by Federal Programm under
contract 14.740.11.0081. The work was also supported  by the DFG
Cluster of Excellence MAP (Munich Centre of Advanced Photonics)
(A.Z.). The work of P.K. is supported in part by the Dynasty
foundation.

\section*{Note Added}
Since the completion of our paper there have been two new papers
on the similar subject. The paper \cite{Rubakov:2010qi} deals with
the proper introduction of the chiral chemical potential in the
theory under consideration. It was argued that $\mu_5$ has to be
coupled to the conserved chiral current, that is the initial
fermionic current has to be modified by the anomalous
contribution. To compare this argument with our approach note that
our additional contribution involving the scalar does the same
job. Indeed, we have argued that vanishing of the canonical
momentum of the scalar implies that the scalar field has a
constant time gradient proportional $\mu_5$. Substituting this
expression for the gradient of the scalar into our additional CS
term we get the expression for the conserved current discussed in
\cite{Rubakov:2010qi}.

In the other paper \cite{Gynther:2010ed} it has been argued that a
nontrivial contribution to the holographic CME in the Lagrangian
without scalars comes from the singular gauge configurations at
the horizon. This statement is still to be clarified. Comparing
this point to our approach with the scalar field we can assume
that the nontrivial effect due to this field may somehow be
related with the singular solutions in \cite{Gynther:2010ed} and
could influence the choice of the gauge. This point certainly
needs further clarification.

It is well-known from the study of the triangle diagram that there
is an ambiguity in the regularization which allows to obtain
either conserved  vector or axial currents. In the standard
situation demanding the conserved vector current we get the
anomaly in the axial current. In the study of the chiral magnetic
effect we can assume that the axial current is conserved instead
of introducing the chiral chemical potential. It is this unusual
viewpoint that amounts to the discussion on the role of the
Bardeen counterterm.

In our model we focus on the scalar Goldstone-Wilczek contribution
to the vector current which is familiar in the chiral theory. This
GW contribution has been overlooked in the previous papers on this
issue. For generality we have presented the different answers
which correspond to different ways to account for the Bardeen
counterterm.

Rubakov calculates the value of current for a differently defined
chemical potential. Namely, his ``axial chemical potential'' is
related to a conserved chiral charge, whereas ours is not. The
difference manifests itself in whether we include the Bardeen
counterterm into the calculation -- it should be taken into
account if we treat the axial chemical potential as a temporal
component of a gauge field.

On the other hand, if the Bardeen counterterm is left out, our
result may be compared with Rubakov's. In that case our result
agrees with both the weak-coupling and with Rubakov's results.
Furthermore, the scalar field contribution in our calculation
corresponds to the anomalous term in the conserved chiral charge
according to Rubakov.

The said ambiguity is that between choice of different models, not
within our model itself; the coefficients in the action of both
our and Rubakov's model are topologically fixed.

Our calculations allow to extract an expression for the axial
current $J_5= -\frac{1}{3}\frac{ N_c\mu}{2\pi^2 B}$ (with the
Bardeen counterterm left out) which is different from one in the
paper by Metlitski and Zhitnitsky\mycite{Metlitski:2005pr}. This
is not surprising since in their paper they consider the
regularization corresponding to the conserved vector current which
is necessary to introduce the standard, not the chiral chemical
potential.


\begin{thebibliography}{99}

\bibitem{Alekseev:1998ds}
  A.~Y.~Alekseev, V.~V.~Cheianov and J.~Fr\"ohlich,
  ``Universality of transport properties in equilibrium, Goldstone theorem  and
  chiral anomaly,''  \textit{Phys. Rev. Lett.} \textbf{81},
  3503–3506 (1998); \texttt{arXiv:cond-mat/9803346}.

\bibitem{Kharzeev:2007jp}
D. E. Kharzeev, L. D. McLerran and H. J. Warringa, ``The effects
of topological charge change in heavy ion collisions: `Event by
event P and CP violation' ", \textit{Nucl. Phys.} \textbf{A803},
227-253 (2008); \texttt{arXiv: 0711.0950[hep-ph]}.

\bibitem{Fukushima:2008xe}
K. Fukushima, D. E. Kharzeev, and H. J. Warringa, ``The Chiral
Magnetic Effect", \textit{Phys. Rev.} \textbf{D78}, 074033 (2008);
\texttt{arXiv: 0808.3382[hep-ph]}


\bibitem{Karch:2007pd}
A. Karch, and A. O'Bannon, ``Metallic AdS/CFT", \textit{JHEP}
\textbf{09}, 024 (2007); \texttt{arXiv: 0705.3870[hep-th]}.

\bibitem{Bergman:2008qv}
O. Bergman, G. Lifschytz and M. Lippert, ``Magnetic properties of
dense holographic QCD", \textit{Phys. Rev.} \textbf{D79}, 105024
(2009); \texttt{arXiv: 0806.0366[hep-th]}.

\bibitem{Lifschytz:2009si}
G. Lifschytz and M. Lippert, ``Anomalous conductivity in
holographic QCD", \textit{Phys. Rev.} \textbf{D80}, 066005 (2009);
\texttt{arXiv: 0904.4772[hep-th]}.

\bibitem{Yee:2009vw}
H.-U. Yee, ``Holographic Chiral Magnetic Conductivity",
\textit{JHEP} \textbf{0911}, 085 (2009); \texttt{arXiv:
0908.4189[hep-th]}.

\bibitem{Metlitski:2005pr}
  M.~A.~Metlitski and A.~R.~Zhitnitsky,
  ``Anomalous axion interactions and topological currents in dense matter,''
  Phys.\ Rev.\  D {\bf 72}, 045011 (2005)
  [arXiv:hep-ph/0505072].

\bibitem{Newman:2005as}
  G.~M.~Newman and D.~T.~Son,
  ``Response of strongly-interacting matter to magnetic field: Some exact
  results,''
  Phys.\ Rev.\  D {\bf 73}, 045006 (2006)
  [arXiv:hep-ph/0510049].

\bibitem{Rebhan:2009vc}
A. Rebhan, A. Schmitt and S. A. Stricker, ``Anomalies and the
chiral magnetic effect in the Sakai-Sugimoto model", \textit{JHEP}
\textbf{1001}, 026 (2010); \texttt{arXiv: 0909.4782[hep-th]}.

\bibitem{Buividovich:2009wi}
P. V. Buividovich, M. N. Chernodub, E. V. Luschevskaya and M. I.
Polikarpov, ``Numerical evidence of chiral magnetic effect in
lattice gauge theory", \textit{Phys.Rev.} \textbf{D80}, 054503
(2009); \texttt{arXiv: 0907.0494[hep-lat]}.

\bibitem{Fukushima:2010fe}
  K.~Fukushima, M.~Ruggieri and R.~Gatto,
  ``Chiral magnetic effect in the PNJL model'',
  arXiv:1003.0047 [hep-ph].

\bibitem{Nam:2009jb}
  S.~i.~Nam,
  Phys.\ Rev.\  D {\bf 80}, 114025 (2009)
  [arXiv:0911.0509 [hep-ph]].

\bibitem{Volovik:2003fe}
  G.~E.~Volovik,
  Int.\ Ser.\ Monogr.\ Phys.\  {\bf 117}, 1 (2006).

\bibitem{Voloshin:2009hr}
\textbf{STAR} collaboration and S. A. Voloshin, ``Experimental
study of local strong parity violation in relativistic nuclear
collisions", \textit{Nucl.Phys.} \textbf{A830}, 377C-384C (2009);
\texttt{arXiv: 0907.2213[nucl-ex]}.

\bibitem{SakaiSugimoto:2005le}
T. Sakai and S. Sugimoto, ``Low energy hadron physics in
holographic QCD", \textit{Prog. Theor. Phys.} \textbf{113},
843-882 (2005); \texttt{arXiv: hep-th/0412141}.

\bibitem{SakaiSugimoto:2005mo}
T. Sakai and S. Sugimoto, ``More on a holographic dual of QCD",
\textit{Prog. Theor. Phys.} \textbf{114}, 1083-1118 (2005);
\texttt{arXiv: hep-th/0507073}.

\bibitem{Karch:2006pv}
  A.~Karch, E.~Katz, D.~T.~Son and M.~A.~Stephanov,
  Phys.\ Rev.\  D {\bf 74}, 015005 (2006)
  [arXiv:hep-ph/0602229].

\bibitem{gw}
  J.~Goldstone and F.~Wilczek,
  ``Fractional Quantum Numbers On Solitons'',
 \textit{Phys. Rev. Lett.}  {\bf 47}, 986 (1981).

\bibitem{Gorsky:2009me}
  A.~S.~Gorsky, V.~I.~Zakharov and A.~R.~Zhitnitsky,
  ``On Classification of QCD defects via holography'',
  Phys.\ Rev.\  D {\bf 79}, 106003 (2009)
  \texttt{arXiv: 0902.1842 [hep-ph]}.\\
A.~Gorsky and V.~Zakharov,
  ``Magnetic strings in Lattice QCD as Nonabelian Vortices'',
  Phys.\ Rev.\  D {\bf 77}, 045017 (2008)
  \texttt{arXiv: 0707.1284 [hep-th]}.
\bibitem{Son:2007ny}
  D.~T.~Son and M.~A.~Stephanov,
  ``Axial anomaly and magnetism of nuclear and quark matter'',
  Phys.\ Rev.\  D {\bf 77}, 014021 (2008)
  \texttt{arXiv: 0710.1084 [hep-ph]}.

\bibitem{Kharzeev:2009pj}
  D.~E.~Kharzeev and H.~J.~Warringa,
  ``Chiral Magnetic conductivity'',
  Phys.\ Rev.\  D {\bf 80}, 034028 (2009)
  \texttt{arXiv:0907.5007 [hep-ph]}.
\bibitem{Kharzeev:2007tn}
  D.~Kharzeev and A.~Zhitnitsky,
  ``Charge separation induced by P-odd bubbles in QCD matter'',
  Nucl.\ Phys.\  A {\bf 797}, 67 (2007)
  \texttt{arXiv: 0706.1026 [hep-ph]}.
\bibitem{gv}
  A.~S.~Gorsky and M.~B.~Voloshin,
  ``Nonperturbative production of multiboson states and quantum
  bubbles'',
  Phys.\ Rev.\  D {\bf 48}, 3843 (1993)
  \texttt{arXiv: hep-ph/9305219}.
\bibitem{sus}
  R.~Kallosh, A.~D.~Linde, D.~A.~Linde and L.~Susskind,
  ``Gravity and global symmetries'',
  Phys.\ Rev.\  D {\bf 52}, 912 (1995)
  \texttt{arXiv: hep-th/9502069}.

\bibitem{Rubakov:2010qi}
  V.~A.~Rubakov,
  arXiv:1005.1888 [hep-ph].
\bibitem{Gynther:2010ed}
  A.~Gynther, K.~Landsteiner, F.~Pena-Benitez and A.~Rebhan,
  arXiv:1005.2587 [hep-th].




\end{thebibliography}
\end{document}